\documentclass[graybox]{svmult}

\usepackage{mathptmx}       
\usepackage{helvet}         
\usepackage{courier}        
\usepackage{type1cm}        
\usepackage{makeidx}         
\usepackage{graphicx}        
\usepackage{multicol}        
\usepackage[bottom]{footmisc}

\usepackage{color,xcolor,amsmath,amssymb,amsfonts}

\newcommand{\IN}{{\rm IN}}
\newcommand{\OUT}{{\rm OUT}}
\newcommand{\WCC}{{\rm WCC}}

\newcommand{\SCC}{{\rm SCC}}
\newcommand{\LSCC}{{\rm LSCC}}

\makeindex             

\begin{document}

\title*{Graph Metrics for Temporal Networks}
\author{Vincenzo Nicosia\inst{1,2} \and John Tang\inst{1} \and Cecilia
  Mascolo\inst{1} \and Mirco Musolesi\inst{3} \and Giovanni
  Russo\inst{4} \and Vito Latora\inst{2,5,6}} \authorrunning{Vincenzo
  Nicosia et al.}
\institute{
  \inst{1} Computer Laboratory, University of Cambridge 15 JJ
  Thomson Avenue, Cambridge CB3 0FD, United Kingdom.
  \and 
  \inst{2}
Laboratorio sui Sistemi Complessi, Scuola Superiore di Catania, Via Valdisavoia 9, 95123 Catania, Italy.
  \and
  \inst{3} School of Computer Science, University of Birmingham,
  Edgbaston, Birmingham B15 2TT, United Kingdom.
  \and
  \inst{4} Dipartimento di Matematica e Informatica,
  Universit\'a di Catania, Via S. Sofia 64, 95123 Catania, Italy.
  \and
  \inst{5} School of Mathematical Sciences, Queen Mary, University of London, E1 4NS, London, United Kingdom
  \and
  \inst{6} Dipartimento di Fisica e Astronomia and INFN, Universit\'a di
  Catania, Via S. Sofia 64, 95123 Catania, Italy.}
%
%
\maketitle

\abstract{Temporal networks, i.e., networks in which the interactions
  among a set of elementary units change over time, can be modelled in
  terms of time-varying graphs, which are time-ordered sequences of
  graphs over a set of nodes. In such graphs, the concepts of node
  adjacency and reachability crucially depend on the exact temporal
  ordering of the links. Consequently, all the concepts and metrics
  proposed and used for the characterisation of static complex
  networks have to be redefined or appropriately extended to
  time-varying graphs, in order to take into account the effects of
  time ordering on causality.  In this chapter we discuss how to
  represent temporal networks and we review the definitions of walks,
  paths, connectedness and connected components valid for graphs in
  which the links fluctuate over time. We then focus on temporal
  node-node distance, and we discuss how to characterise link
  persistence and the temporal small-world behaviour in this class of
  networks. Finally, we discuss the extension of classic centrality
  measures, including closeness, betweenness and spectral centrality,
  to the case of time-varying graphs, and we review the work on temporal motifs
  analysis and the definition of modularity for temporal graphs.}

\abstract*{ }

\section{Introduction}

Whenever a system consists of many single units interacting through a
certain kind of relationship, it becomes natural to represent it as a
\textit{graph}, where each \textit{node} of the graph stands for one
of the elementary units of the system and interactions between
different units are symbolised by \textit{edges}. If two nodes are
connected by an edge they are said to be \textit{adjacent}. According
to the nature of the units and to the characteristics of adjacency
relationship connecting them, it is possible to construct different
kind of graphs, such as friendship graphs -- where nodes are people
and edges connect two people who are friends--, functional brain
networks -- where nodes are regions of the brain and edges represent
the correlation or causality of their activity--, communication graphs
-- where nodes are terminals of a communication systems, such as
mobile phones or email boxes, and edges indicate the exchange of a
message from a terminal to another, just to make some examples. Thanks
to the availability of large data sets collected through modern
digital technologies, in the last decade or so there has been an
increasing interest towards the study of the structural properties of
graph representations of real systems, mainly spurred by the
observation that graphs constructed from different social, biological
and technological systems show surprising structural similarities and
are characterised by non--trivial properties. In a word, they are
\textit{complex} networks. Independently of the peculiar nature and
function of the original systems, the corresponding graphs are usually
\textit{small--worlds}, i.e., they show high local cohesion and, at
the same time, extremely small node-node distance~\cite{Watts1998};
the distribution of the number of neighbours of a node (its
\textit{degree}) is often heterogeneous, and decays as a power--law
(i.e., they are \textit{scale--free}~\cite{Albert1999}); they are
locally organised as tightly-knit groups of nodes (called
\textit{communities}), which are in turn loosely interconnected to
each other~\cite{Newman2006}.  The concepts, metrics, methods,
algorithms and models proposed so far to study the structure of real
networks has led to the formation of theoretical framework known as
\textit{complex network
  theory}~\cite{Barabasi2002rev,Newman2003rev,Boccaletti2006}.

However, the relationships among the units of a real networked system
(e.g., node adjacency) are rarely persistent over time. In many cases,
the static interpretation of node adjacency is just an oversimplifying
approximation: contacts among individuals in a social network last
only for a finite interval and are often intermittent and
recurrent~\cite{Holme2005,Clauset2007,Kossinets2008}; different
intellectual tasks are usually associated to different brain activity
patterns~\cite{Valencia2008,Fallani2008}; communication between agents
in a telecommunication system are typically bursty and fluctuate over
time~\cite{Barabasi2005,Stehle2010,Miritello2011}; transportation
networks show fluctuations in their microscopic organisation, despite
the stability of their global structural
properties~\cite{Gautreau2009}. Consequently, whenever we deal with a
networked system that evolves over time, the concept of adjacency
needs to be appropriately redefined. The extension of node adjacency
to the case of time-evolving systems has lately led to the definition
of \textit{temporal graphs} (sometimes also called
\textit{time-varying graphs} or \textit{dynamic graphs}), in which
\textit{time} is considered as another dimension of the system and is
included in the same definition of the graph. Since most of the
metrics to characterise the structure of a graph, including graph
connectedness and components, distance between nodes, the different
definitions of centrality etc., are ultimately based on node
adjacency, they need to be appropriately redefined or extended in
order to take into account of the presence, frequency and duration of
edges at different times. In general, the temporal dimension adds
richness and complexity to the graph representation of a system, and
demands for more powerful and advanced tools which allow to exploit
the information on temporal correlations and causality. Recently,
Holme and Saram\"aki have published a comprehensive review which
presents the available metrics for the characterisation of temporal
networks~\cite{Holme2012}. A description of some potential
applications of these metrics can be found
in~\cite{TLSNMML13:applications}.

This chapter presents the basic concepts for the analysis of
time-evolving networked systems, and introduces all the fundamental
metrics for the characterisation of time-varying graphs. In
Section~\ref{sec:tvg} we briefly discuss different approaches to
encode some temporal information into static graphs and we introduce a
formal definition of time-varying graph. In Section~\ref{sec:connect}
we examine how reachability and connectedness are affected by
time-evolving adjacency relationships and we introduce the definitions
of node and graph temporal components, showing the intimate
connections between the problem of finding temporal connected
components and the maximal-clique problem in static graphs. In
Section~\ref{sec:dist} we focus on the concepts of temporal distance,
efficiency and temporal clustering, and we discuss the temporal
small-world phenomenon. In Section~\ref{sec:centr} we present the
extensions of betweenness, closeness and spectral centrality to
time-varying graphs. Finally, in Section~\ref{sec:meso} we report on
the definition of temporal motifs and on the extension of the
modularity function to time-varying graphs.

\section{Representing Temporal Networks} 
\label{sec:tvg}

\begin{figure}[t]
  \centering \includegraphics[width=4in]{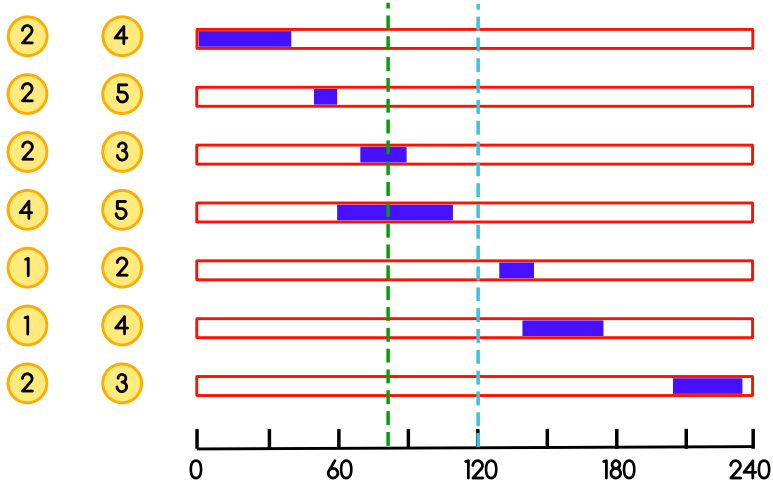}
  \caption{The set of contacts registered among five nodes within an
    observation period of $4$ hours. Blue bars indicate the duration
    of each contact. The two dashed lines correspond to two
    instantaneous cuts of the contact set, respectively for $t=80$min
    (green) and $t=120$min (cyan).}
  \label{fig:contacts}
\end{figure}

From a mathematical point of view a networked time-evolving system
consists of a set $\mathcal{C}$ of \textit{contacts} registered among
a set of nodes $\mathcal{N}=\{1,2,\ldots,N\}$ during an
\textit{observation interval} $[0,T]$~\cite{Tang2009,Pan2011}. A
\textit{contact} between two nodes $i,j \in \mathcal{N}$ is
represented by a quadruplet $c=(i,j,t,\delta t)$, where $0\le t\le T$
is the time at which the contact started and $\delta t$ is its
duration, expressed in appropriate temporal units. As we stated above
the relationship between $i$ and $j$ is usually not persistent (it
could represent the co-location at a place, the transmission of a
message, the temporal correlation between two areas of the brain
etc.), so that in general we will observe more than one contact
between $i$ and $j$ in the interval $[0,T]$.  In
Figure~\ref{fig:contacts} we report an example of a set of seven
contacts observed between a set of $N=5$ nodes within an interval of
$T=240$ minutes. The contacts in the figure are considered
\textit{symmetric}, i.e., $(i,j,t,\delta t) = (j,i,t,\delta t)$, even
if in general this in not the case. Each contact is represented by a
pair of nodes and a blue bar indicating the start and duration of the
contact. Notice that in this example, which is indeed
representative of many social and communication systems, the typical
overlap between contacts is relatively small with respect to the
length of the observation interval.

Before explaining how the temporal information about such a set of
contacts can be represented by means of a time-varying graph, we first
review some simple approaches to deal with time-evolving systems based
on static graphs, and we discuss why they are inadequate for analyzing
time-evolving systems.

\subsection{Aggregated Static Graphs}

The classic approach to represent
networked systems evolving over time consists in constructing a
single \textit{aggregated static graph}, in which all the contacts
between each pair of nodes are flattened in a single edge. An
aggregated graph can be represented by an \textit{adjacency matrix}
$A=\{a_{ij}\} \in \mathbb{R}^{N\times N}$, in which the entry
$a_{ij}=1$ if at least one contact $(i,j,\cdot,\cdot)$ has been
registered in $[0,T]$ between $i$ and $j$, and $a_{ij}=0$
otherwise. If the relationship between any pair of nodes $i$ and $j$
is symmetric, such as in the case of co-location or collaboration graphs,
also the corresponding adjacency matrix is symmetric, i.e., $a_{ij} =
a_{ji}\quad \forall i,j\in\mathcal{N}$. Conversely, whenever a
directionality is implied, for instance when the contact is a phone
call from $i$ to $j$ or represents goods transferred from $i$ to $j$,
the adjacency matrix is in general non-symmetric.

Representing a time-evolving system by means of an adjacency matrix,
i.e., an unweighted graph, is usually a severe
oversimplification: the information about the
number, frequency and duration of contacts between two nodes $i$ and
$j$ is flattened down into a binary digit (i.e., $a_{ij} =1$ if there
is at least one contact, of any duration, between $i$ and $j$, while
$a_{ij}=0$ otherwise). In general, a binary adjacency information does
not take into account the heterogeneity observed in real systems. Just
to make an example, both the number of phone calls made by a single
node during a certain time interval the duration of a call between
two nodes exhibit large fluctuations, and their distribution is well
approximated by a
power--law~\cite{Barabasi2005,Miritello2011,Karsai2011}. This means
that assigning the same weight to all the relationships can lead to
misleading conclusions. This problem can be partially solved by
constructing a \textit{weighted aggregated graph}, in which the edge
connecting $i$ to $j$ is assigned a weight $w_{ij}$ proportional to
the number of contacts observed, their duration, their
frequency or a combination of the these three dimensions.

\begin{figure}[t]
  \centering 
  \includegraphics[width=4.5in]{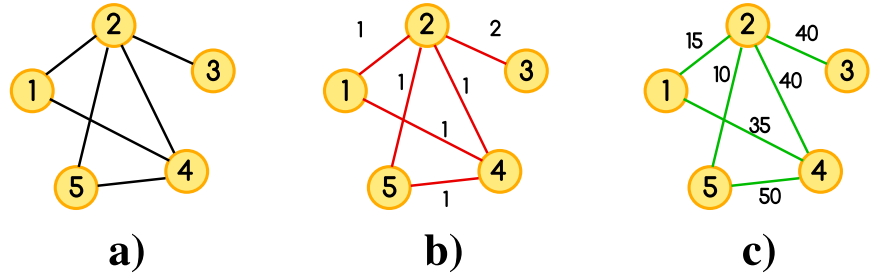}
  \caption{Three different aggregated static graphs obtained from the
    set of contacts in Figure~\ref{fig:contacts}. a) The unweighted
    aggregated graph; b) the weighted aggregated graph where each
    weight $w_{ij}$ between node $i$ and node $j$ corresponds to the
    number of contacts observed; c) the weighted aggregated graph
    where each weight $w_{ij}$ is the total duration of all the
    contacts between $i$ and $j$.}
    \label{fig:aggregated}
\end{figure}

In Figure~\ref{fig:aggregated} we show three different aggregated
static graphs corresponding to the same set of contacts reported in
Figure~\ref{fig:contacts}. The leftmost graph (Panel a) is the
unweighted aggregated graph; in the middle graph (Panel b) the weights
correspond to the number of contacts observed between the nodes; in
the rightmost one (Panel c) the weight of each edge $w_{ij}$ is equal
to the sum of the duration of all the contacts between $i$ and
$j$. However, both unweighted and weighted aggregated graphs fail to
capture the temporal characteristics of the original system. In fact,
by considering all the links as always available and persistent over
time, the number of walks and paths between two nodes is
overestimated, while the effective distance between two nodes is
instead systematically underestimated. For instance, in all the three
aggregated representations, node $2$ and node $4$ are connected by an
edge, but their interaction is limited just to the beginning of the
observation period, so that these nodes cannot directly communicate
for most of the time.

Despite not being powerful enough to represent networks in which
the temporal aspects are crucial, static aggregated graphs and the
metrics proposed for their analysis still constitute an invaluable
framework to investigate the structure and function of systems in
which the topological characteristics are more relevant than the
temporal ones.  After all, most of the classic examples of complex
networks, including the graph of the Internet at Autonomous Systems
level~\cite{Vazquez2002}, co-authorship
networks~\cite{Newman2001,Newman2001a}, the graph of the World Wide
Web~\cite{Albert1999,Broder2000} and functional brain
networks~\cite{Bullmore2009} have been obtained so far by aggregating
all the contacts observed among a certain number of nodes within a
given temporal interval, and the analysis of their structure has
provided new insights about the organisation of different complex
systems.

\subsection{Time-varying Graphs}

The natural way to work out a graph representation that can properly
take into account all the temporal correlations of a set of contacts
consists into including time as an additional dimension of the
graph~\cite{Kempe2002}. We notice that a set of contacts implicitly
defines a graph for each instant $t$, made by the set of edges
corresponding to all the contacts $(\cdot, \cdot, t_i, \delta t_{i})$
such that $t_i \le t \le t_i + \delta t_{i}$. In the example shown in
Figure~\ref{fig:contacts}, if we consider $t=80$min, corresponding to
the dashed green line in the figure, the graph constructed from
contacts active at that time contains only two edges, namely $(2,3)$
and $(4,5)$. However, we notice that the graph corresponding to
$t=120$min (the dashed cyan line in the figure) is an empty graph,
since no contact is active at that time. For practical reasons, and
especially when contacts are instantaneous, i.e., $\delta t \rightarrow
0$, it is convenient to consider a finite \textit{time-window} $[t,
  t+\Delta t]$ and to construct a graph by creating an edge between
all pairs of nodes for which there is at least a contact which
overlaps with the interval $[t, t+\Delta t]$. A generic contact
$(\cdot, \cdot, \tau_i, \delta \tau_i)$ overlaps with $[t, t+\Delta
  t]$ if it satisfies at least one of the three following conditions:

\begin{eqnarray}
  t \le \tau_i < t + \Delta t\\
  t \le \tau_i + \delta \tau_i < t + \Delta t\\
  \tau_i < t \quad \wedge \quad \tau_i + \delta \tau_i > t + \Delta t
\end{eqnarray}

A graph $G_t$ obtained by aggregating all the contacts
appearing in a given interval $[t, t+\Delta t]$ represents the
state of the system in that interval, i.e., it is a \textit{snapshot}
which captures the configuration of the links at that particular time
interval. If we consider a sequence of successive non-overlapping
time-windows $\{[t_1, t_1 + \Delta t_1], [t_2, t_2 + \Delta t_2],
[t_3, t_3 + \Delta t_3], \ldots, [t_{M}, t_M + \Delta t_M]\}$ then we
obtain a \textit{time-varying graph}, which is the simplest graph
representation of a set of contacts that takes into account their
duration and their temporal ordering~\cite{Tang2009,Kostakos2008}.
A time-varying graph is an ordered sequence of $M$ graphs
$\{G_1,G_2,\ldots,G_M \}$ defined over $N$ nodes, where each graph
$G_m$ in the sequence represents the state of the network, i.e., the
configuration of links, in the time-window $[t_m, t_m+\Delta t_m],
\quad m=1,\ldots,M$. In this notation, the quantity $t_M + \Delta t_M
-t_1$ is the temporal length of the observation period.  In general
the graphs in the sequence can correspond to any ordered sequence of
times such that $t_1< t_1 + \Delta t_1 = t_2< t_2 + \Delta t_2 = t3 <
\ldots <t_M+\Delta t_M$~\cite{Grindrod2011}. In the following we
assume, without loss of generality, that $t_1 = 0$ and $t_M=T$ and, at
the same time, that the sequence of snapshots is uniformly distributed
over time, i.e., $t_{m+1} =t_m + \Delta t, ~\forall
m=1,\ldots,M-1$~\cite{Tang2009}. In compact notation, we denote the
graph sequence forming a time-varying graph as $\mathcal{ G}\equiv
\mathcal{G}_{[0,T]}$.  Each graph in the sequence can be either
undirected or directed, according to the kind of relationship
represented by contacts. Consequently, the time-varying graph
$\mathcal{G}$ is fully described by means of a time-dependent
adjacency matrix $A(t_{m}),~~m=1,\ldots,M$, where $a_{ij}(t_m)$ are
the entries of the adjacency matrix of the graph at time $t_m$, which
is in general a non-symmetric matrix.

\begin{figure}[t]
  \centering 
  \includegraphics[width=4.5in]{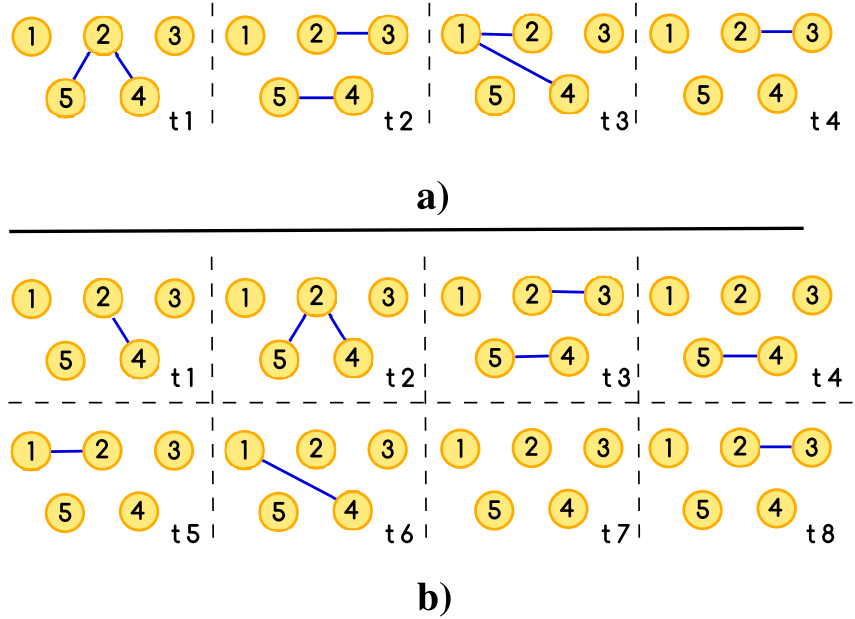}
  \caption{Two time-varying graphs corresponding to the set of
    contacts in Figure~\ref{fig:contacts}. Panel a): the four snapshot
    obtained setting $\Delta t=60$min; Panel b): the eight snapshots
    of the time-varying graph constructed using $\Delta t=30$min. The
    smaller the size of the time-window, the higher the probability
    that a snapshot contains no edges (this happens for the snapshot
    $t7$ in Panel b).}
    \label{fig:time_scale}
\end{figure}

Notice that, by tuning the size of the time-window used to construct
each snapshot, it is possible to obtain different representations of
the system at different temporal scales. In
Figure~\ref{fig:time_scale} we present two time-varying graphs
obtained from the same set of contacts in Fig.~\ref{fig:contacts} but
using two different lengths for the time-window. The graph on the top
panel is constructed by setting $\Delta t = 60$min, and consists of
four snapshots, while the graph on the bottom panel corresponds to a
time-window of $\Delta t=30$min and has eight snapshots. It is usually
preferable to set the size of the time-window to the maximum temporal
resolution available. For instance, if the duration of contacts is
measured with an accuracy of one second (such as in the case of email
communications or phone calls), it makes sense to construct
time-varying graphs using a time-window $\Delta t = 1s$.

In the limit case when $\Delta t \rightarrow 0$, we obtain an
infinite sequence of graphs, where each graph corresponds to the
configuration of contacts at a given instant $t$. This
sequence of graphs might include a certain number of empty graphs,
corresponding to periods in which no contacts are registered.  On the
contrary, if we set $\Delta t = T$, the time-varying graph degenerates
into the corresponding unweighted aggregated graph, where all the
temporal information is lost.

\section{Reachability, Connectedness and Components}
\label{sec:connect}

In a static graph the \textit{first neighbours} of a node $i$ are the
nodes to which $i$ is connected by an edge, i.e., nodes $j$ such that
$a_{ij} = 1$. We say that the neighbours of $i$ are \textit{directly
  reachable} from $i$. If $k$ is a neighbour of $j$ and $j$
is in turn a neighbour of $i$, then the node $k$ is indirectly
reachable from $i$, i.e., by following first the edge connecting $i$ to
$j$ and then the one which connects $j$ to $k$.  In general, the
direct and indirect reachability of nodes is important to characterise
the global structure of a network and to investigate the dynamics of
processes occurring over it. For instance, if node $i$ has got a
contagious disease, then there is a high probability that the disease
will sooner or later be transmitted to the nodes that are directly
reachable from $i$ (its first neighbours). However, if a disease
starts from a node $i$, also the nodes which are not directly
connected to $i$ but are still indirectly reachable from $i$ have a
finite probability to get the disease through a chain of
transmissions.

The reachability between nodes is related to the concept of
\textit{walk}. In a static graph a walk is defined as an ordered
sequence of $\ell$ edges $\{a_{i_0,i_1}, a_{i_1,i_2},\ldots,
a_{i_{\ell-1}, i_\ell}\}$ such that $a_{i_k, i_{k+1}}=1,
k=0,1,\ldots,\ell-1$. The \textit{length} of a walk is equal to the
number of edges traversed by the walk. We say that the node $j$ is
\textit{reachable} from $i$ if there exists a walk which starts at $i$
and ends up at $j$. If each vertex in a walk is traversed exactly
once, then the walk is called a \textit{simple walk} or a
\textit{path}.  For instance, in the graph shown in
Figure~\ref{fig:aggregated} the sequence of nodes $[2,4,5,2,1,4]$ is a
walk of length $5$ which starts at node $2$ and ends at node $4$,
while the sequence $[3,2,5]$ is a path of length $2$ going from node
$3$ to node $5$ passing by node $2$. In a static graph the length of
the shortest path connecting two nodes is called \textit{geodesic
  distance}.

Since the definitions of walk and path depend on the adjacency of
nodes, and given that node adjacency is a function of time in
time-varying graphs, an appropriate extension of these concepts is
necessary in order to define node reachability and components in
time-varying graphs.

\subsection{Time-respecting Walks and Paths}

In a time-varying graph, a {\em temporal walk} from node $i$ to node
$j$ is defined as a sequence of $L$ edges $[(n_{r_0},n_{r_1}),
  (n_{r_1},n_{r_2}), \ldots, (n_{r_{L-1}},n_{r_L})]$, with $n_{r_0}
\equiv i$, $n_{r_L} \equiv j$, and an increasing sequence of times
$t_{r_1} < t_{r_2} < \ldots < t_{r_L}$ such that $a_{n_{r_{l-1}} ,
  n_{r_l}}(t_{r_l}) \neq 0~~
l=1,\ldots,L$~\cite{Tang2009,Grindrod2011}.
A {\em path} (also called {\em temporal path}) of a time-varying graph
is a walk for which each node is visited at most once. For instance,
in the time-varying graph of Figure~\ref{fig:time_scale}a, the
sequence of edges $[(5,2), (2,1)]$ together with the sequence of times
$t_1,t_3$ is a temporal path of the graph. This path starts at node
$5$ at time $t_1$ and arrives at node $1$ at time $t_3$.  Notice that
the aggregated static graph flattens down most of the information
about temporal reachability. In fact, if we look at the static
aggregated graph corresponding to this time-varying graph (shown in
Figure~\ref{fig:aggregated}a), there are different paths going from
node $1$ to node $5$ and viceversa; however, if we look at the
time-varying graphs of Figure~\ref{fig:time_scale} we notice that in
both of them there is no temporal path connecting node $1$ to node
$5$. The reason is that node $5$ could be reached from node $1$ only
by passing through either node $2$ or node $4$, but node $1$ actually
is connected to both these nodes \textit{after} they have been in contact to node
$5$.

\subsection{Temporal Connectedness and Node Components}

The concept of connectedness is fundamental in complex network
theory.  A message, a piece of information or a disease can be
transferred from one node to all the other nodes to which it is
connected, but will never be conveyed to nodes that are disconnected
from it. For this reason, the study of node connectedness and node
components is the very basic tool to investigate the structure of a
graph. 

In a static undirected graph two nodes are said to be connected if
there exists a path between them. In this particular case
connectedness is an equivalence relation: it is \textit{reflexive}
(i.e., $i$ is connected to itself), \textit{symmetric} (i.e., if $i$
is connected to $j$ then $j$ is connected to $i$) and
\textit{transitive} (i.e., if $i$ is connected to $j$ and $j$ is
connected to $k$, then $i$ is also connected to $k$). Instead, in a
directed graph, due to the directionality of the edges, symmetry is
broken and the existence of a path from $i$ to $j$ does not guarantee
that a path from $j$ to $i$ does indeed exist. For this reason, the
notions of strong and weak connectedness are introduced. In a word,
two nodes $i$ and $j$ of a static directed graph are said to be
\textit{strongly connected} if there exist a path from $i$ to $j$ and
a path from $j$ to $i$, while they are \textit{weakly connected} if
there exists a path connecting them in the underlying undirected
graph, i.e., in the static graph obtained from the original by
discarding edge directions.

Starting from the definitions of temporal walk and path, it is
possible to define temporal connectedness (in a weak and in a strong
sense) for pairs of nodes in a time-varying graph. A node $i$ of a
time-varying graph $\mathcal{G}_{[0, T]}$ is {\em temporally
  connected} to a node $j$ if there exists a temporal path
going from $i$ to $j$ in $[0,T]$. Due to the temporal ordering of edges, this
relation is trivially not symmetric, so that if $i$ is temporally
connected to $j$, in general $j$ can be either temporally connected or
disconnected to $i$. Two nodes $i$ and $j$ of a time-varying graph are
{\em strongly connected} if $i$ is temporally connected to $j$ and
also $j$ is temporally connected to $i$.

Temporal strong connectedness is a reflexive and symmetric relation,
so that if $i$ is strongly connected to $j$, then $j$ is strongly
connected to $i$. However, strong connectedness still lacks
transitivity, and, therefore, it is not an equivalence relation. In
fact, if $i$ and $j$ are strongly connected and $j$ and $l$ are
strongly connected, nothing can be said, in general, about the
connectedness of $i$ and $l$. For instance, in the time-varying graph
shown in Figure~\ref{fig:time_scale}a, node $5$ and $2$ are strongly
connected and also $2$ and $1$ are strongly connected, but nodes $5$
and $1$ are not strongly connected because, as we have already
explained above, there exists no temporal path that connects node~$1$
to node~$5$.

Similarly to the case of static directed graphs, it is possible to
define weak connectedness among nodes. Given a time-varying graph
$\mathcal{G}$, we consider the underlying undirected time-varying
graph $\mathcal{G}^u$, which is obtained from $\mathcal{G}$ by
discarding the directionality of the links of all the graphs
$\{G_m\}$, while retaining their time ordering.  Two nodes $i$ and $j$
of a time-varying graph are {\em weakly connected} if $i$ is
temporally connected to $j$ and also $j$ is temporally connected to
$i$ in the underlying undirected time-varying graph $\mathcal{G}^u$.
Also weak connectedness is a reflexive and symmetric relation, but not
transitive.

It is worth noting that strong and weak connectedness propagate over
different time scales. In fact, if we consider two time-varying graphs
obtained from the same set of contacts by using two different
time-windows, such as for instance $\Delta t_1$ for the graph
$\mathcal{G}_1$ and $\Delta t_2 > \Delta t_1$ for $\mathcal{G}_2$ (as
in the two time-varying graphs of Figure~\ref{fig:time_scale}), then
it is easy to prove that if $i$ and $j$ are strongly connected in
$\mathcal{G}_1$ then they are also strongly connected in
$\mathcal{G}_2$. The contrary is trivially not true. Thanks to this
property, strong and weak connectedness in time-varying graphs are
consistent with the corresponding definitions given for static
graphs. In fact, as a limiting case, if two nodes are strongly
(weakly) connected in a time-varying graph, then they are also
strongly (weakly) connected in the corresponding aggregated static
graph, which is the degenerate time-varying graph obtained by setting
$\Delta t = T$.

By using reachability, strong and weak connectedness, different
definitions of node components can be derived. For instance, the {\em
  temporal out-component of node $i$} (resp. \textit{in-component)},
denoted as $\OUT_T(i)$ (resp. $\IN_T(i)$), is the set of vertices
which can be reached from $i$ (resp. from which $i$ can be reached) in
the time-varying graph $\mathcal{G}$. Similarly the {\em temporal
  strongly connected component of a node $i$} (resp. \textit{weakly
  connected component}), denoted as $\SCC_T(i)$ (resp. $\WCC_{T}(i)$),
is the set of vertices from which vertex $i$ can be reached, and which
can be reached from $i$, in the time-varying graph $\mathcal{G}$
(resp. in the underlying undirected time-varying graph
$\mathcal{G}^u$).

In general, temporal node components have quite heterogeneous
composition and sizes, and reveal interesting details about the
structure of the graph. For instance, the out-component of node $3$ in
the two graphs of Figure~\ref{fig:time_scale} contains only nodes $1$,
$2$, $4$ and node $3$ itself, since there is no way for node $3$ to
reach node $5$. Conversely, in the corresponding aggregated graphs (as
shown in Figure~\ref{fig:aggregated}) the out-components of all the
nodes are identical and contain all the nodes of the graph. 

The importance of temporal node components has been pointed out in
Ref.~\cite{Nicosia2012}, which reports the results of temporal
component analysys on time-varying graphs obtained from three
different data sets. The authors compared the size of node temporal
in- and out-components in these time-varying graphs with the size of
the giant component of the corresponding aggregated graphs, and they
found that in general temporal node components are much smaller than
the giant component of the aggregated graph, and exhibit a high
variability in time. This is another example of the fact that
time-varying graphs are able to provide additional information that is
not captured by aggregated graphs.

\subsection{Graph Components and Affine Graphs}

\begin{figure}[t]
  \centering \includegraphics[width=1.4in]{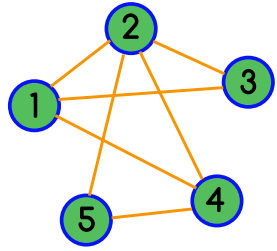}
  \caption{Affine graph associated to the time-varying graph of
    Figure~\ref{fig:time_scale}a. Pairs of strongly connected nodes
    are linked by an edge, and the cliques of the affine graph
    correspond to the strongly connected components of the associated
    time-varying graph.}
  \label{fig:affine}
\end{figure}

Differently from the case of directed static graphs, it is not
possible to define the strongly (weakly) connected components of a
time-varying graph starting from the definition of connectedness for
pairs of nodes. Formally, this is due to the fact that strong and weak
connectedness are not equivalence relations. For this reason, the
following definition of strongly connected component of a time-varying
graph has been proposed~\cite{Nicosia2012}: a set of nodes of a
time-varying graph $\mathcal{G}$ is a temporal strongly connected
component of $\mathcal{G}$ if each node of the set is strongly
connected to all the other nodes in the set. A similar definition
exists for weakly connected components. These definitions enforce
transitivity but have the drawback that in order to find the strongly
connected components of a time-varying graph, it is necessary to check
the connectedness between all pairs of nodes in the graph. In
Ref.~\cite{Nicosia2012} it has been shown that the problem of finding
the strongly connected components of a time-varying graph is
equivalent to the well-known problem of finding the maximal-cliques of
an opportunely constructed static graph.  We define such graph as the
\textit{affine graph} corresponding to the time-varying graph. The
affine graph $G_{\mathcal{G}}$ is defined as the graph having the same
nodes as the time-varying graph $\mathcal{G}$, and such that two nodes
$i$ and $j$ are linked in $G_{\mathcal{G}}$ if $i$ and $j$ are
strongly connected in $\mathcal{G}$. In Fig.~\ref{fig:affine} we
report the affine graph corresponding to the time varying graph shown
in Fig.~\ref{fig:time_scale}a.
In this graph, node $1$ is directly connected to nodes $\{2,3,4\}$,
since it is temporally strongly connected to them in the time-varying
graph. Similarly, node $2$ is connected to nodes $\{1, 3, 4, 5\}$,
node $3$ is connected to $\{1,2\}$, node $4$ is connected to $\{1, 2,
5\}$ and node $5$ is connected to $\{2,4\}$. Hence, the affine graph
$G_{\mathcal{G}}$ has only 7 of the 10 possible links, each link
representing strong connectedness between two nodes. By construction,
a clique of the affine graph $G_{\mathcal{G}}$ contains only nodes
which are strongly connected to each other, so that the
\textit{maximal-cliques} of the affine graph, i.e., all the cliques
which are not contained in any other clique, are temporal strongly
connected components ($\SCC_T$) of $\mathcal{G}$.  Similarly, all the
\textit{maximum-cliques} of the affine graph $G_{\mathcal{G}}$, i.e.,
its largest maximal-cliques, are the largest temporal strongly
connected components ($\LSCC_T$) of $\mathcal{G}$. 

We notice that the problem of finding a partition of $\mathcal{G}$
that contains the minimum number of disjoint strongly connected
components is equivalent to the well--known problem of finding a
partition of the corresponding affine graph $G_{\mathcal{G}}$ in the
smallest number of disjoint
maximal-cliques~\cite{Karp1972}. Unfortunately, this problem is known
to be NP--complete, and in practice can be exactly solved only for
small graphs. In the case of the affine graph reported in
Figure~\ref{fig:affine}, it is possible to check by hand that there
are only three possible partitions of $G_{\mathcal{G}}$ into
maximal-cliques, namely
\begin{enumerate}
\item $\{1, 2, 3\} \bigcup \{4, 5\}$
\item $\{1, 2, 4\} \bigcup \{3\} \bigcup \{5\}$
\item $\{2, 4, 5\} \bigcup \{1, 3\}$
\end{enumerate}
The second partition contains two isolated nodes, which are indeed
degenerated maximal-cliques.  Therefore, the original time-varying
graph admits only two different partitions into a minimal number of
non-degenerated strongly connected components, namely into two
components containing at least two nodes each. This is a quite
different picture from that we obtain using static aggregated
graphs. In fact, all the static aggregated representations of the same
time-varying graph (see Figure~\ref{fig:aggregated}) are composed by
just one strongly connected component which includes all the nodes.

We notice that in general the largest temporal strongly connected
components of a time-varying graph can be much smaller than the giant
connected component of the corresponding aggregated graph. For
instance, in Ref.~\cite{Nicosia2012} the authors performed temporal
component analysis on time-varying graphs constructed from three
different time-stamped data sets (i.e. the MIT Reality Mining project,
co-location at INFOCOM 2006 and Facebook communication), and they
found that despite the giant connected component of the corresponding
aggregated graphs usually included almost all the nodes in the
network, the maximal cliques of the affine graphs were indeed much
smaller. Particularly interesting was the case of the Facebook
communication data set: the giant connected components of the
aggregated graphs contained from $10^4$ to $10^5$ nodes, while the
largest temporal strongly connected components counted around one
hundred nodes at most. Disregarding such discrepancies could result in
misleading conclusions. For example, the potential number of
individuals infected by a disease which spreads through the system is
in the order of tens if we correctly take into account temporal
correlations, but could be erroneously estimated to be thousand times
larger if one considers the aggregated graph.

\section{Distance, Efficiency and Temporal Clustering}
\label{sec:dist}

One of the most relevant properties of static complex networks is that
they exhibit, on average, a surprising small geodesic distance between
any pairs of nodes, where the geodesic distance between $i$ and $j$ is
defined as the length of the shortest path connecting them. The
average geodesic distance is important to characterise how fast (for
example, in terms of number of hops), a message can be transmitted
from a node to another in the network; therefore, it is related to the
overall efficiency of communication among nodes. Having a small
average geodesic distance (where the average is computed over all the
pairs of connected nodes) is a desirable property when one wants to
spread a message throughout the network; conversely, small geodesic
distance becomes a problem if we want to control the propagation of a
disease.

\subsection{Temporal Distance and Efficiency}

When we consider time-varying graphs, the temporal dimension is an
essential element of the system, so that the concept of geodesic
distance cannot be limited to the number of \textit{hops} separating
two nodes but should also take into account the temporal ordering of
links. As a matter of fact, any path in a time-varying graph is
characterised by two different lengths: {\em a)} a \textit{topological
  length}, measured as the number of edges traversed by the path, and
{\em b)} a \textit{temporal length} or duration, measured as the time
interval between the first and the last contact in the path. 

Both the topological and the temporal lengths of a path usually depend
on the time at which the path starts. Consider for instance two of the
paths connecting node $5$ to node $3$ in Figure~\ref{fig:time_scale}a.
The first path starts from $5$ at snapshot $t_1$ which traverses the
edge $(5,2)$ at $(t_1)$ and arrives at node $3$ following the edge
$(2,3)$ at time $(t_2)$. The second one starts from $5$ at time $t_2$
and arrives at node $3$ at time $t_4$, after traversing the edges
$(5,4)$ at $t_2$, $(4,1)$ and $(1,2)$ at $t_3$ and finally $(2,3)$ at
$t_4$. The first path has a topological length equal to $2$ and a
temporal length of two snapshots (e.g., $2$ hours), while the second
path has a topological length equal to $4$ and a temporal length equal
to three snapshots (e.g., $3$ hours).

The \textit{temporal shortest path} from node $i$ to node $j$ is
defined as the temporal path connecting $i$ to $j$ which has minimum
temporal length. Similarly, the \textit{temporal distance} $d_{i,j}$
between $i$ and $j$ is the temporal length of the temporal shortest
path from $i$ to $j$. In the example discussed above, the temporal
shortest path connecting node $5$ to node $3$ is the one starting at
$t_1$ and having temporal length equal to two snapshots.

The natural extension of the average geodesic distance to the case of
time-varying graphs is the \textit{characteristic temporal path
  length}~\cite{Tang2009,Tang2010a}, which is defined as the average
temporal distance over all pairs of nodes in the graph:

\begin{equation}
  L = \frac{1}{N(N-1)}\sum_{ij}d_{ij}
\end{equation}

It is also possible to define the \textit{temporal diameter} of a
graph as the largest temporal distance between any pair of nodes:

\begin{equation}
  D = max_{ij} d_{ij}
\end{equation}

However, in real time-varying graphs it is quite common to have many
pairs of temporally disconnected nodes. The problem is that if a node
$j$ is not temporally reachable from $i$, then $d_{ij}=\infty$, and
the characteristic temporal path length diverges. In order to avoid
such divergence, the \textit{temporal global
  efficiency}~\cite{Tang2009,Tang2010a} of a time-varying graph has
been defined as follows:

\begin{equation}
  \mathcal{E} = \frac{1}{N(N-1)}\sum_{ij}\frac{1}{d_{ij}}
\end{equation}

The temporal global efficiency is the straightforward generalisation
of the global efficiency already defined for static
graphs~\cite{Latora2001}, and has been successfully employed to study
and quantify the robustness of temporal graphs (see for instance
Ref.~\cite{SLMBZ12:evaluating} and~\cite{TLSNMML13:applications} in
this book).

\subsection{Edge Persistence and Clustering}

The characteristic temporal path length and the temporal efficiency
provide a quantitative representation of the global structure of a
graph in terms of the average temporal distance among any pair of
nodes. However, in time-varying systems contacts are usually
\textit{bursty}, meaning that the distribution of the time between two
contacts has a heavy tail, and \textit{persistent}, i.e., if two nodes
are connected at a time $t$, there is a non-negligible probability
that they will still be connected at time $t + \Delta t$. This
characteristic can be quantified in the following way. If we consider
a node $i$ and two adjacent snapshots of a time-varying graph,
respectively starting at time $t_{m}$ and $t_{m+1}= t_m+\Delta t$, we
can define the \textit{topological overlap} of the neighbourhood of
$i$ in $[t_m, t_{m+1}]$ as:

\begin{equation}
C_i(t_m,t_{m+1}) = \frac{\sum_j a_{ij}(t_m)
  a_{ij}(t_{m+1})}{\sqrt{\left[\sum_j a_{ij}(t_m)\right] \left[\sum_j
      a_{ij}(t_{m+1})\right]}}
\end{equation}

and the \textit{average topological overlap} of the neighbourhood of
node $i$ as the average of $C_i(t_m,t_{m+1})$ over all possible
subsequent temporal snapshot, i.e.:

\begin{equation}
  C_i = \frac{1}{M-1}\sum_{m=1}^{M-1} C_i(t_m,t_{m+1})
\end{equation}

The average topological overlap of a node $i$ is a natural extension
of the concept of local clustering coefficient which includes temporal
information. In fact, while in a static graph the local clustering
coefficient of a node measures the probability that its neighbours are
in turn connected by an edge, the average topological overlap
estimates the probability that an edge from $i$ to one of its
neighbours $j$ persists across two consecutive time-windows. In a
word, it is a measure of the \textit{temporal clustering} of edges,
i.e., of their tendency to persist across multiple windows.  The
average of $C_i$ computed over all the nodes in the network, namely
the quantity:
\begin{equation}
C =\frac{1}{N}\sum_i C_i
\end{equation}
is called \textit{temporal-correlation coefficient}~\cite{Tang2009},
and is a measure of the overall average probability for an edge to
persist across two consecutive snapshots. Notice that $C=1$ if and
only if all the snapshots of the time-varying graphs have exactly the
same configuration of edges, while it is equal to zero if none of the
edges is ever observed in two subsequent snapshots.

\begin{table}[!htb]
\centering
\begin{tabular}{c|cccccc}
           & $C$ & $C^{rand}$   & $L$                & $L^{rand}$        & $E$    & $E^{rand}$ \cr  
  \hline
  \hline 
$\alpha$  & 0.44& 0.18 (0.03)        &    3.9     & 4.2    & 0.50    & 0.48     \cr
$\beta$   & 0.40& 0.17 (0.002)        &    6.0    & 3.6    & 0.41    & 0.45     \cr
$\gamma$  & 0.48& 0.13 (0.003)        &    12.2    & 8.7    & 0.39    & 0.37     \cr
$\delta$  & 0.44& 0.17 (0.003)         &    2.2   & 2.4     & 0.57    & 0.56     \cr 
\hline
\hline
d1  & 0.80 & 0.44 (0.01)    &   8.84     &  6.00    & 0.192   &  0.209     \cr
d2  & 0.78 & 0.35 (0.01)   &   5.04     &  4.01    & 0.293   &  0.298     \cr
d3  & 0.81 & 0.38 (0.01)   &   9.06      &  6.76    & 0.134   &  0.141     \cr
d4  & 0.83 & 0.39 (0.01)   &   21.42     &  15.55   & 0.019   &  0.028     \cr
\hline
\hline
Mar & 0.044 & 0.007 (0.0002)  &   456               & 451  & 0.000183 & 0.000210 \cr
Jun & 0.046 & 0.006 (0.0002)  &   380               & 361  & 0.000047 & 0.000057 \cr
Sep & 0.046 & 0.006 (0.0002)  &   414               & 415  & 0.000058 & 0.000074 \cr
Dec & 0.049 & 0.006 (0.0002)  &   403               & 395  & 0.000047 & 0.000059 \cr 
  \hline
  \hline 
\end{tabular}
\caption{Temporal-correlation, characteristic temporal path length and
  efficiency for brain functional networks (in four different
  frequency bands)~\cite{Fallani2008}, for the social interaction
  networks of INFOCOM'06 (time periods between 1pm and 2:30pm, four
  different days)\cite{Scott2009}, and for messages over Facebook
  online social network (four different months of year
  2007)\cite{Wilson2009}.  Results are compared with the averages
  measured over $1000$ time-varying graphs obtained by reshuffling the
  sequences of snapshots.  The values in parenthesis next to
  $C_{rand}$ are the respective standard deviations.  The values of
  $L$ and $L^{rand}$ are computed considering only the connected pairs
  of nodes. Table adapted from~\cite{Tang2009}}
\label{tab:table}
\end{table}

In Ref.~\cite{Tang2009} the authors considered time-varying graphs
constructed from three data sets, namely functional brain
networks~\cite{Fallani2008}, the co-location at INFOCOM
2006~\cite{Scott2009} and personal messages exchanged among Facebook
users~\cite{Wilson2009}. They compared the characteristic temporal path
length and the temporal correlation coefficient of these temporal
graphs with those obtained from the same data sets by reshuffling the
sequence of snapshots. Notice that by reshuffling the snapshots one
destroys all the existing temporal correlations while retaining the
average connectivity of each node and the configuration of edges in
the corresponding aggregated graph. They showed that the original
time-varying graphs usually exhibit both a relatively smaller
characteristic temporal path length and a relatively higher temporal
correlation coefficient, when compared with those measured on
reshuffled sequences of snapshots.  This finding is the temporal
analogous to the small--world effect observed in static complex
networks, and has consequently been named \textit{small--world
  behaviour in time-varying graphs}. Table~\ref{tab:table} reports the
results obtained for the three different data sets.

\section{Betweenness, Closeness and Spectral Centrality}
\label{sec:centr}

The structural properties of a complex network usually reveal
important information about its dynamics and function. This is
particularly true if we take into account the relationship between the
position occupied by a node in a static graph and the \textit{role}
played by the node for the evolution of a dynamic process. For
instance, not all nodes have the same impact on the transmission of a
disease (or the spreading of a rumour) over a network: intuitively, the
nodes having a higher number of neighbours should contribute much more
to the spreading than nodes having few connections. However, if we
perform a deeper analysis, we observe that not just the number of
edges is important to identify good spreaders, since also the actual
organisation of these edges has an impact on the speed of the
spreading process. In fact, nodes mediating a large number of shortest paths
are indeed those that contribute the most to the transmission of
diseases and information over a network. The identification of nodes
that play a central role, i.e., nodes having high \textit{centrality},
has been a quite active research field in complex network
theory. Here we review some standard centrality measures and their
extension to the case of time-varying graphs.

\subsection{Betweenness and Closeness Centrality}

Two basic centrality measures based on shortest paths are
\textit{betweenness} centrality and \textit{closeness} centrality. The
betweenness centrality of a node $i$ in a static graph is defined as
follows:
\begin{equation}
  C_i^{B} = \sum_{j\in V} \sum_{\stackrel{k \in V} {k \neq j}}
  \frac{\sigma_{jk}(i)}{\sigma_{jk}}
    \label{eq:betweenness}
\end{equation}
where $\sigma_{jk}$ is the number of shortest paths from node $j$ to
node $k$, while $\sigma_{jk}(i)$ is the number of such shortest paths
that pass through the node $i$. The higher the number of shortest
paths passing through $i$, the higher the value of $C_i^B$.
Betweenness centrality can be also defined for single edges, by
counting the fraction of shortest paths between any pair of nodes to
which a given edge participate.

A simple way to extend betweenness centrality to time-varying graphs
consists in counting the fraction of temporal shortest paths that
traverse a given node. The formula would be exactly the same as
Eq.~\ref{eq:betweenness}, with the only difference that $\sigma_{jk}$
and $\sigma_{jk}(i)$ will be, respectively, the total number of
temporal shortest paths between $j$ and $k$ and the number of those
paths which make use of node $i$.

Sometimes it can be important to take into account not only the number
of temporal shortest paths which pass through a node, but also the
length of time for which a node along the shortest path retains a
message before forwarding it to the next node~\cite{Tang2010}. For
example, let us consider the simple case of nodes $i$ and $j$ being
connected by just one shortest path which consists of the two edges
$(i,k)_{t_\ell}$ and $(k,j)_{t_m}$. This means that the edge
connecting $i$ to $k$ appears at time $t_\ell$, while the edge
connecting $k$ to $j$ appears at time $t_m$. Since the path through
$k$ is the only way for $i$ to temporally reach $j$, then we would say
that $k$ plays an important mediatory role and is "central" for
communication between $i$ and $j$. Nevertheless, the vulnerability of
node $k$ heavily depends on the interval $[t_\ell, t_m]$: the longer
this temporal interval, the higher the probability that a message
forwarded to $k$ is lost if $k$ is removed from the network. In order
to take into account the effect of waiting times, the \textit{temporal
  betweenness centrality}~\cite{Tang2010} of the node $i$ at time $t_m$ is defined as:
\begin{equation}
  C_i^B(t_m) = \frac{1}{(N-1)(N-2)}\sum_{j\neq i}\sum_{\stackrel{k\neq
      j}{k \neq i}}\frac{U(i,t_m,j,k)}{\sigma_{jk}}
\end{equation}

where $\sigma_{jk}$ is the number of temporal shortest path from $j$
to $k$, and $U(i,t_m,j,k)$ is the number of temporal shortest paths
from $i$ to $j$ in which node $i$ is traversed from the path in the
snapshot $t_m$ or in a previous snapshot $t'<t_m$, so that the next
edge of the same path will be available at a later snapshot $t'' >
t_m$. The \textit{average temporal betweenness} of node $i$ is defined
as the average of $C_i^{B}(t_m)$ over all the snapshots:
\begin{equation}
  C_i^{B} = \frac{1}{M} \sum_{m} C_i^{B}(t_m)
\end{equation}

The closeness centrality of a node $i$ is a measure of how close $i$
is to any other node in the network. It can be measured as the inverse
of the average distance from $i$ to any other node in the network:
\begin{equation}
  C_i^C = \frac{N-1}{\sum_j d_{ij}}
\end{equation}
where $d_{ij}$ is the distance between $i$ and $j$ in a static
graph. The \textit{temporal closeness centrality} is defined in an
analogous way, the only difference being that for time-varying graphs
$d_{ij}$ denotes the length of the temporal shortest path from $i$ to
$j$.

As shown in Ref.~\cite{Tang2010} and elsewhere in this
book~\cite{TLSNMML13:applications}, temporal closeness and betweenness
centrality have proven useful to identify key spreaders and temporal
mediators in corporate communication networks. In particular, it was
found that traders indeed played an important mediatory role in
time-varying graphs constructed from the ENRON email communication
data set, being consistently ranked among the first ones both for
temporal betweenness and for temporal closeness centrality. This
result is qualitatively and substantially different from the one
obtained by computing betweenness and closeness centrality in the
corresponding aggregated graph, where the most central nodes are the
people who interacted with the most number of other people, i.e., a
secretary and a managing director. This apparently unimportant
discrepancy between the centrality rankings actually turns out to be
fundamental for the spreading of information (or diseases) throughout
the system. In fact, simulation reported in Ref.~\cite{Tang2010}
confirmed that when a spreading process is initiated at the nodes
having the highest temporal closeness centality the number of other
nodes reached by the spreading was higher and the time needed to reach
them was smaller than in the case in which the spreading starts at
nodes having higher static closeness centrality.

\subsection{Spectral Centrality and Communicability}

The total number of shortest paths passing through a node is not
always the best way of measuring its centrality, especially because
the shortest paths are not always the most relevant for a process. For
instance, a disease could propagate through any path (not just through
the shortest ones), and the rumours usually follow walks which are
much longer (and somehow less efficient) than shortest
paths. Consequently, other definitions of centrality exist which take
into account walks instead of shortest paths. The classic example for
static graphs is represented by the so called \textit{Katz
  centrality}~\cite{Newman2010}. The basic ides is that the propensity
for node $i$ to communicate with node $j$ can be quantified by
counting how many walks of length $\ell=1,2,3,\ldots$ lead from $i$ to
$j$. The importance of a walk of length $\ell=1$ (i.e., the direct
edge $(i,j)$) is higher than that of a walk of length $\ell=2$, which
in turn is higher than that of a walk of length $\ell=3$ and so
forth. For this reason, it makes sense to appropriately rescale the
contribution of longer walks. The original proposal consisted into
scaling walks of length $\ell$ by a factor $\alpha^{\ell}$, where
$\alpha$ is an appropriately chosen real value. We notice that the
element $a_{ij}^\ell$ of the $\ell^{th}$ power of the adjacency matrix
corresponds to the number of existing walks of length $\ell$ between
$i$ and $j$. Consequently, the entry $s_{ij}$ of the matrix sum $S =
I+\alpha A+ \alpha^2 A^2+\ldots$ measures the propensity of $i$ to
interact with $j$ (notice that $I$ is the $N\times N$ identity
matrix). It is possible to prove that the sum $S$ converges to
$(I-\alpha A)^{-1}$ if $\alpha<\rho(A)$, where $\rho(A)$ is the
spectral radius of the adjacency matrix. In this case, the Katz
centrality of node $i$ is measured as the sum of the $i^{th}$ row of
$S$:
\begin{equation}
  C_i^{K} = \sum_{j} \left[(I-\alpha A)^{-1}\right]_{ij}
\end{equation}

Katz centrality can be extended to the case of time-varying graphs by
using a similar reasoning~\cite{Grindrod2011}. We notice that each
entry of the product of the adjacency matrices corresponding to an
increasing sequence of $\ell$ snapshots $[t_{r_1}, t_{r_2}, \ldots,
  t_{r_\ell}]$ represents the number of temporal walks in which the
first edge belongs to the snapshot $t_{r_1}$, the second edge to
$t_{r_2}$ and so on. So, in order to count all the possible temporal
walks of any length we should sum over all possible products of the
form:

\begin{equation}
  \alpha^{k}A(t_{r_1})A(t_{r_2})\cdots A(t_{r_k}), \quad t_{r_1} \le
  t_{r_2}\le \cdots \le t_{r_k}
\end{equation}
for any value of the length $k$. It is possible to prove that if
$\alpha < \min_{t_m}{\rho(A(t_m))}$ then the sum of all these products
can be expressed as:

\begin{equation}
  Q \equiv \left[I-\alpha A(t_0)\right]^{-1}\left[I-\alpha
      A(t_1)\right]^{-1}\cdots \left[I-\alpha A(t_m)\right]^{-1}
\end{equation}
The matrix $Q$ is called \textit{communicability matrix}. Starting
from this matrix one can define the \textit{broadcast centrality}:
\begin{equation}
  C_i^{Broad} = \sum_{j}Q_{ij}
\end{equation}
which quantifies how well node $i$ can reach all the other nodes in
the time-varying graph, and the \textit{receive communicability}:
\begin{equation}
  C_i^{Recv} = \sum_{j}Q_{ji}
\end{equation}
which is an estimation of how well node $i$ can be reached from any
other node in the network. In Ref.~\cite{Grindrod2011} it has been
found that broadcast and receive communicability can be useful to spot
the most influential spreaders in different time-evolving
communication networks.

\section{Meso-scale Structures}
\label{sec:meso}

Real static networks differ from random graphs in many ways. In fact,
together with heterogeneous distributions of node properties
(e.g. degree and centrality) and with specific global characteristics
(e.g. high average clustering and small average path length), complex
networks show a non-trivial organisation of subsets of their nodes and
exhibit a variety of meso-scale structures, including \textit{motifs}
and \textit{communities}. The characterisation of the abundance of
specific motifs has helped to explain why biological and technological
networks are relatively resilient to
failures~\cite{Alon2002,Alon2004}, while the analysis of communities
has revealed that there exists a tight relationship between structure
of a network and its functioning~\cite{Fortunato2009}.
In the following we discuss how motifs analysis can be performed also
in time-varying graphs and we present the extension of the modularity
(a function for measuring the quality of a partition in communities,
defined for static graphs by Newman in Ref.~\cite{Newman2004}) to
temporal communities.

\subsection{Temporal Motifs}

In static graphs a \textit{motif} is defined as a class of isomorphic
subgraphs. We recall that two graphs $G'$ and $G''$, having adjacency
matrices $A'$ and $A''$, are \textit{isomorphic} if there exists a
permutation of the labels of the nodes of $A'$ such that, after the
permutation, $A'\equiv A''$. A permutation is represented by a matrix
$P$ that has the effect of swapping the rows and columns of the matrix
to which it is applied. If $A'$ and $A''$ are isomorphic, then there
exists a permutation matrix $P$ such that:
\begin{equation}
  P^{-1} A' P = A''
\end{equation}
If two graphs are isomorphic then they are topologically equivalent,
i.e., the arrangement of the edges in the two graphs is exactly the
same, up to an appropriate relabelling of the nodes. Consequently, a
motif can be thought as the typical representative of a class of
subgraphs sharing the same arrangement of edges.  It has been shown
that in real networks, especially in biological ones, motifs are not
uniformly distributed and some motifs are over-represented while
others are rare~\cite{Alon2002,Alon2004}.

As for all the metrics described so far, the extension of
\textit{motifs} to time-varying graphs has to take into account time
in a meaningful way.  In a recent paper Kovanen et
al.~\cite{Kovanen2011} propose an extension of motifs to
time-varying graphs, based on the definition of $\Delta
\tau$-adjacency and $\Delta \tau$-connectedness of
contacts\footnote{In order to avoid confusion with the size $\Delta
  t$ of the time-window used to define the temporal snapshot of a
  time-varying graph, here we preferred to use $\Delta \tau$ instead
  of the original $\Delta t$ proposed by the authors
  of~\cite{Kovanen2011}. Also, notice that the the authors use to call
  \textit{events} what we have called here \textit{contacts}.}. For
practical reasons, the authors made the simplifying assumption that
each node can be involved in no more than one contact at a time. This
assumption is in general too restrictive, but it could be valid in
some cases, e.g. when the contacts represent phone calls or when the
duration $\delta t$ of a contact is so small that the probability for
a node to have two simultaneous contacts is negligible.

We say that two contacts $c_a=(i,j,t_a, \delta t_a)$ and $c_b=(k,\ell,
t_b, \delta t_b)$ are $\Delta \tau$-adjacent if they have at least one
node in common and the time difference between the end of the first
contact and the beginning of the second one is no longer than $\Delta
\tau$. We assume, without loss of generality, that $t_a < t_b$, so
that $c_a$ and $c_b$ are $\Delta \tau$-adjacent if $0 \le t_b - t_a -
\delta t_a \le \Delta\tau$. We say that an ordered pair of contacts
$(c_a, c_b)$ is \textit{feasible} if $t_a<t_b$.
Notice that
$\Delta\tau$-adjacency is defined only for the subset of feasible
pairs of contacts. Two contacts $c_a$ and $c_b$ are
$\Delta\tau$-connected if there exists a sequence of $m$ contacts
$S=\{c_a=c_{n_0},c_{n_1}, c_{n_2}, \ldots, c_{n_m}=c_b \}$ such that
each pair of consecutive contacts in $S$ is feasible and
$\Delta\tau$-adjacent.  From $\Delta\tau$-connectedness, we derive the
definition of \textit{connected temporal subgraph}, which is a set of
contacts such that all feasible pairs of contacts in the set are
$\Delta \tau$-connected. 

\begin{figure}
  \centering \includegraphics[width=4.5in]{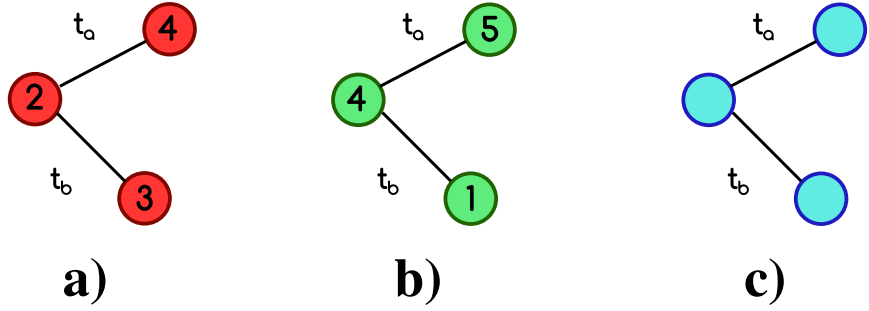}
  \caption{Connected temporal subgraphs and motifs. The subgraph in
    panel a) is not a valid temporal subgraph, because node $2$ is
    involved in another contact after the contact with node $4$ at
    time $t_a$ and before the contact with node $3$ at time
    $t_b$. Conversely, the subgraph in Panel b) is valid. Panel c)
    shows the motif associated to the valid subgraph in Panel b).}
  \label{fig:motifs}
\end{figure}

For the definition of temporal motifs, we restrict ourselves to the
subset of \textit{valid temporal subgraphs}. A temporal subgraph is
considered valid if all the $\Delta\tau$-adjacent contacts of the
nodes in the subgraph are consecutive. This means that if a node $j$
appears in the pair of $\Delta\tau$-adjacent contacts $c_a=(i,j,t_a,
\delta t_a)$ and $c_b=(j,k, t_b, \delta t_b)$ of the subgraph, then
does not exist any contact $c_x = (j,k, t_x, \delta t_x)$ such that
$t_a < t_x < t_b$. In Figure~\ref{fig:motifs} we report two temporal
subgraphs of three nodes obtained from the set of contacts in
Figure~\ref{fig:contacts} considering $\Delta\tau=30$min. The subgraph
of Panel a) corresponds to the sequence of contacts $S_1=\{c_1 =
(2,4,0,40), c_3 = (2,3,70,20)\}$, and is not valid because node $2$ is
involved in another contact, namely $c_2 = (2,5,50,10)$ after $c_1$
and before $c_3$. Conversely, the connected temporal subgraph reported
in Panel b) and corresponding to the pair of contacts $S_2 =
\{c_4=(4,5,60,50),c_6=(1,4,140,35)\}$, is valid, since node $4$ is not
involved in any contact between $c_4$ and $c_6$.

A \textit{temporal motif} is a class of isomorphic valid temporal
subgraphs, where two temporal subgraphs are considered isomorphic if
they are topologically similar (i.e., the organisation of the links in
the subgraph is equivalent up to an appropriate relabelling of the
nodes) and represent the same temporal pattern, i.e., the order of the
sequence of contacts is the same. The typical element of the temporal
motif corresponding to the graph reported in Figure~\ref{fig:motifs}b
is shown in Figure~\ref{fig:motifs}c, where the labels on the edges of
the graph correspond to the ordering of contacts. In
Ref.~\cite{Kovanen2011} the authors report also an algorithm to
discover temporal motifs, and discuss the problems connected with the
estimation of the significance of motifs.

\subsection{Temporal Communities and Modularity}

The identification of communities, i.e., groups of tightly connected
nodes, has allowed to reveal the richness of static graphs and has
helped to understand their organisation and function.
The simplest way to partition a graph is by dividing it into a set of
$\mathcal{M}$ non-overlapping groups, so that each node of the graph
is assigned to one of the $\mathcal{M}$ communities. The quality of a
non-overlapping partition in communities can be measured by the
\textit{modularity function}. This function, originally proposed by
Newman in Ref.~\cite{Newman2004}, estimates the difference between the
fraction of edges among nodes belonging to the same community and the
expected fraction of such edges in a null-model graph with no
communities. More formally, it is defined as follows:
\begin{equation}
  Q=\frac{1}{2K}\sum_{ij} (a_{ij} - P_{ij})\delta(c_i, c_j)
\end{equation}
where $a_{ij}$ are the elements of the adjacency matrix, $P_{ij}$ is
the expected number of edges between $i$ and $j$ in the null-model
graph, $c_i$ is the community to which node $i$ belongs,
$\delta(c_i,c_j)=1$ if and only if node $i$ and node $j$ belong to the
same community and $K$ is the total number of edges in the graph. The
simplest null-model is represented by a configuration model graph,
where all the nodes of the graph have the same degree as in the real
graph but edges are placed at random.  In this case the modularity
function reads:
\begin{equation}
  Q=\frac{1}{2K}\sum_{ij} \left(a_{ij} -
  \frac{k_ik_j}{2K}\right)\delta(c_i, c_j)
\end{equation}
where $k_i$ is the degree of node $i$.  Different extensions of the
modularity function have been proposed for directed graphs, weighted
graphs and graphs with overlapping
communities~\cite{Newman2004b,Arenas2007,Leicht2008,Nicosia2009}.

The extension of modularity for time-varying graphs is based on an
interesting result valid for the modularity of static graphs and
presented in Ref.~\cite{Lambiotte2008}, which connects the modularity
function with the dynamics of a random walk over the graph. The
authors of Ref.~\cite{Lambiotte2008} show that the modularity function
can be considered a particular case of a class of functions that
measure the dynamical stability of a partition $\mathcal{P}$, where
the stability of $\mathcal{P}$ at time $t$ is defined as:
\begin{equation}
  R(t) = \sum_{\mathcal{C} \in \mathcal{P}} P(\mathcal{C}, t) -
  P(\mathcal{C}, \infty)
\end{equation}
Supposing that the random walk has reached the stationary
state\footnote{It is possible to prove that a random walk on a graph
  always converges towards a stationary state, independently of the
  initial condition, if the adjacency matrix of the graph is
  primitive, which is the case for the vast majority of real graphs.},
then $P(\mathcal{C},t)$ is the probability that a random walker which
starts from a node in the community $\mathcal{C}$ is found in a node
of $\mathcal{C}$ after time $t$. Similarly, $P(\mathcal{C}, \infty)$
is the probability that a random walker that started from a node in
$\mathcal{C}$ is found in $\mathcal{C}$ after an infinite number of
steps; when the walk has reached the stationary state, this
corresponds to the probability that two independent random walks are
found in $\mathcal{C}$ at the same time. It is possible to show that
if we consider a discrete-time random walk, in which a walker jumps
from a node to another at equally-spaced time-steps of length
$\Delta\tau$, then the stability of a partition at one step
${R}(t=\Delta\tau)$ is identical to the modularity function.

In Ref.~\cite{Mucha2010} Mucha et. al propose an extension of
modularity to multi-slice graphs which exploits the connection between
the modularity function and the stability of a random walk on the
multi-slice graph. Indeed, a time-varying graph can be considered a
multi-slice graph if we connect each node of a snapshot with the other
instances of itself in neighbouring snapshots by means of
\textit{multi-slice edges}. For brevity, we give here the definition
of the modularity function for multi-slice graphs, which corresponds
to the stability at one step of a random walk on the multi-slice
graph, but we omit the derivation of the formula\footnote{The
  interested reader can find the derivation of
  Eq.~\ref{eq:modul_multi} in Ref.~\cite{Mucha2010} and in the
  Supplemental Information of the same paper.}. The modularity for
multi-slice graphs reads:
\begin{equation}
  Q_{multi}=\frac{1}{2\mu}\sum_{ijsr}\left[\left(a_{ij}(s) - \gamma_s
    \frac{k_i(s)k_j(s)}{2m_{s}}\right)\delta_{sr} +
    C_{jsr}\delta_{ij}\right]\delta(g_{is},g_{jr})
  \label{eq:modul_multi}
\end{equation}
The indices $i$ and $j$ are used for nodes while the indices $r$ and
$s$ indicate different slices. Here $a_{ij}(s)$ are the elements of
the adjacency matrix of slice $s$, $k_{i}(s)$ represents the degree of
node $i$ in slice $s$ (i.e., the number of neighbours to with $i$ is
connected in that slice) and $m_s=\frac{1}{2}\sum_{i}k_{i}(s)$ is the
total number of edges in the slice $s$. The term $C_{jsr}$ is the
weight of the link that connects node $j$ in slice $s$ to itself in
slice $r$, and $\gamma_s$ is a resolution parameter. The terms
$\delta_{ij}$ an $\delta_{rs}$ indicate the Kronecker function and
$\delta(g_{is},g_{jr})$ is equal to $1$ only if node $i$ in slice $s$
and node $j$ in slice $r$ belong to the same community. The definition
looks a bit complicated but it is essentially composed of two
parts. The term in parentheses represents the standard modularity for
the graph at slice $s$ (the only difference being the resolution
parameter $\gamma_s$), while the term $C_{jsr}$ accounts for
inter-slice connections. Once we have defined modularity for
multi-slice networks, the search for the best partition can be
performed by using one of the standard methods for modularity
optimisation~\cite{Fortunato2009}.

This definition of modularity is quite general, works well for any
kind of multi-slice network, and is also applicable to assess the
quality of a partition of a time-varying graph, which can be
considered a multi-slice network. However, when using
Eq.~\ref{eq:modul_multi} one should take into account that in order to
derive it as the stability of a random walk on the graph, the edges
connecting different slices have to be undirected\footnote{This is
  required to ensure the existence of a stationary state for the
  Laplacian dynamics on the graph.}. Consequently, this definition of
modularity in invariant under inversion of the sequence of slices
which, in the particular case of time-varying graphs, implies
invariance under time inversion. This means that
Eq.~\ref{eq:modul_multi} gives the same result on the time-varying
graph $\mathcal{G}_{[0,T]}$ and on the graph $\mathcal{G}_{[T, 0]}$ in
which the sequence of time-windows is given in the opposite order.

In general, invariance under time inversion is not a desirable property
for a metric used to characterise the structure of time-varying
graphs, because most of the interesting characteristics of
time-evolving systems, including temporal correlation of links and
reachability, are due to the asymmetry introduced by the \textit{arrow
  of time}. Time-invariance disregards this asymmetry completely,
washing out most of the richness of time-varying
systems. Consequently, we believe that the definition of appropriate
metrics for the evaluation of community structures in time-varying
graphs is still an open field of investigation.

\section{Final Remarks}

The description of temporal networks in terms of time-varying graphs
and the analysis of their structural properties is still in its
infancy but has already produced many encouraging results, showing
that complex networks theory is a quite flexible and promising
framework for the characterisation of different real systems. There
are still some open problems to be tackled, such as the definition of
appropriate methods to detect temporal communities and the
construction of analytical methods to assess how the structure of a
time-varying graph can affect the dynamics of processes occurring over
it, including spreading, synchronisation and evolutionary
games. However, even if the community has not yet converged towards a
unified notation and a fully consistent set of definitions and
approaches is still lacking, the metrics and concepts devised so far
for time-varying graphs constitute a valid and consistent alternative
to the standard methods for the study of time-evolving systems, and
will certainly represent a fundamental contribution to our
understanding of complex systems in general.

\begin{acknowledgement}
This work was funded in part through EPSRC Project MOLTEN
(EP/I017321/1) and the EU LASAGNE Project, Contract No.318132 (STREP).
\end{acknowledgement}

\bibliographystyle{plain} \bibliography{nicosia_book2012}

\end{document}